\documentclass[3p,authoryear]{elsarticle}
\usepackage{amsmath}
\usepackage{amsthm}
\usepackage{comment}
\usepackage{url}
\usepackage{txfonts}
\usepackage{graphicx}  
\usepackage[utf8]{inputenc}
\usepackage[capitalise]{cleveref}
\usepackage{xcolor}
\usepackage{lscape}
\usepackage{color}
\usepackage{soul}
\usepackage{rotating}
\usepackage{appendix}
\usepackage{subcaption}

\usepackage{multirow}
\usepackage{here}

\usepackage{threeparttable}

\usepackage{bigstrut}
\usepackage[normalem]{ulem}

\newcommand{\add}[1]{\textcolor{black}{#1}}

\begin{document}

\begin{frontmatter}

\title{
Delay propagation patterns in Japan's domestic air transport network
}

\address[TSE]{Department of Transdisciplinary Science and Engineering, Tokyo Institute of Technology, Tokyo 152-8550, Japan}

\author[TSE]{Kashin Sugishita\corref{cor}}\ead{sugishita.k.aa@m.titech.ac.jp}\cortext[cor]{Corresponding author}
\author[TSE]{Kazuki Arisawa}
\author[TSE]{Shinya Hanaoka}

\begin{abstract}\quad 
We experience air traffic delays every day, but are there any recurrent patterns in these delays?
In this study, we investigate the recurrence of delay propagation patterns in Japan's domestic air transport network in 2019 by integrating delay causality networks and temporal network analysis.
Additionally, we examine characteristics unique to delay propagation by comparing delay causality networks with corresponding randomized networks generated by a directed configuration model.
As a result, we found that the structure of the delay propagation patterns can be classified into several groups. 
The identified groups exhibit statistically significant differences in total delay time and average out-degree, with different airports playing central roles in spreading delays.
The results also suggest that some delay propagation patterns are particularly prominent during specific times of the year, which could be influenced by Japan's seasonal and geographical factors.
Moreover, we discovered that specific network motifs appear significantly more (or less) frequently in delay causality networks than their corresponding randomized counterparts.
This characteristic is particularly pronounced in groups with more significant delays.
These results suggest that delays propagate following specific directional patterns, which could significantly contribute to predicting air traffic delays.
We expect the present study to trigger further research on recurrent and non-recurrent natures of air traffic delay propagation.
\end{abstract} 


\end{frontmatter}

\section{Introduction}
\subsection{Delay propagation in air transport networks}

Air transport system is an essential part of global transportation, which connects people and goods worldwide and enables trade, tourism, and cultural exchange.
Even though the COVID-19 pandemic impacted the aviation industry significantly, passenger traffic has been on the recovery \citep{IATA2023air}.
Air transport demand is projected to increase at an average annual growth rate of 4.3\% over the next 20 years \citep{ICAO2023future}, and the number of passengers is expected to reach 8.2 billion by 2037 \citep{IATA2018passenger}.

The accelerating increase in air traffic demand, coupled with the limitations of airspace capacity, is exerting immense pressure on the global air transport system, leading to a surge in flight delays \citep{czerny2010airport, pyrgiotis2013modelling}.
In fact, on average, about 30,000 flights a day are delayed worldwide \citep{forbes2023delay}.
This frequency of air traffic delays causes a myriad of negative impacts.  
Firstly, flight delays can lead to dissatisfaction among passengers \citep{britto2012impact, kevsel2019risk}, potentially resulting in diminished trust and loyalty towards airlines \citep{vlachos2014drivers}.
The delays also lead to financial losses for airlines. 
Airlines face financial burdens resulting from the necessary reallocation of resources \citep{abdelghany2004model} and the escalated passenger compensation provided to maintain the demand \citep{hu2016integrated}.
The delays can also harm the market share and reputations of airlines and airports \citep{peterson2013economic}.
From an environmental perspective, delays can intensify concerns about fuel consumption and subsequent greenhouse gas emissions \citep{rebollo2014characterization, sternberg2017review}.

The factors contributing to air traffic delays can be classified into occurrence and propagation factors. 
The occurrence factors include extreme weather, air carrier issues, and air traffic control \citep{BTS2021delay}.   
On the other hand, the propagation factors include aircraft rotation \citep{lan2006planning, zou2014flight} and transit of passengers and crews \citep{beatty1999preliminary, kafle2016modeling}. 
Because of these factors, a minor delay at one airport can potentially trigger large-scale propagation across the entire network.  

Here, one question arises: \add{how can we classify daily delay propagation into distinct patterns, and what characterizes each of these patterns?}
To address this question, we examine recurrence in the structure of daily delay propagation in Japan's domestic air transport network throughout 2019. 
Recurrence implies that the structure of the network returns to one that is similar to the network in the past \citep{masuda2019detecting, sugishita2021recurrence}. 
Understanding the recurrence of the delay propagation would significantly contribute to predicting air traffic delays.
Note that the reason for choosing the data of 2019 is to analyze the recurrence in delay propagation under normal operational conditions before the COVID-19 pandemic.
In addition, we are interested in airline-specific delay propagation patterns. 
Therefore, we investigate delay propagation patterns not only for the entire domestic network but also for the two major full-service carriers in Japan, All Nippon Airways (ANA) and Japan Airlines (JAL).

\subsection{Related studies}

Significant progress has been made in the research on air traffic delay. 
\cite{malone1995dynamic} proposed Approximate Network Delay (AND) model for three airports network.
\cite{pyrgiotis2013modelling} further developed the AND model and analyzed the network with 34 airports in the US.
\cite{ahmadbeygi2008analysis} studied the relationship between planned schedules and delay propagation for two major US airlines.
\cite{fleurquin2013systemic} proposed a delay propagation model that considers aircraft rotation, transit of passengers and crews, and airport congestion.
\cite{campanelli2014modeling} further developed a model that includes slot reallocation and swapping and analyzed delay propagation in Europe.
\cite{hao2014new} investigated the impact of three New York airports on delays throughout the network.
\cite{kafle2016modeling} analyzed delay propagation in the US air transport network with an analytical-econometric approach. 
\cite{wu2018modeling} proposed an Airport-Sector Network Delays model to estimate the flight delay, considering both airports and airspace capacities. 
\cite{wu2019modelling} developed a delay propagation model using a Bayesian network in a delay-tree framework.
\cite{brueckner2021airline} conducted a theoretical and empirical analysis of airlines' choice of schedule buffers.
\cite{tan2021exploratory} developed an econometric model to quantify
delay propagation and analyzed the Chinese air transport network.
These studies have contributed significantly to the understanding of delay propagation.
However, these methods are limited in describing delay only in terms of explicit interactions and spatiotemporal correlations \citep{zeng2022research}.

Researchers have recently started applying causal inference to understand air traffic delays, even though the number of studies is still limited.
Typically, these studies perform causal inference among time series of delays at airports and then construct and analyze the so-called delay causality networks, in which nodes are airports, and edges are delay causalities.
For the causal inference, Granger causality \citep{granger1969investigating} has been most commonly used \citep{zanin2017network, du2018delay}. 
Still, other techniques have also been used, such as nonlinear Granger causality \citep{jia2022delay} and convergent cross mapping \citep{guo2022detecting}.
These studies have contributed significantly to understanding the structure of delay propagation. 
However, there is one aspect that has not been sufficiently examined in the previous studies - the recurrence of delay propagation patterns. 

In the last two decades, studies on temporal networks, in which network structure changes over time, have been rapidly advanced \citep{holme2012temporal, holme2019temporal, masuda2020guidance}.
Real-world network systems often change their structure over time, and temporal network analysis is required to understand the dynamics of such systems.
There has been extensive research on various properties of temporal networks such as burstiness \citep{lambiotte2013burstiness, moinet2015burstiness}, temporal centrality \citep{pan2011path, kim2012temporal}, and temporal community structure \citep{mucha2010community, fortunato2016community}.
Recently, methods for analyzing evolution patterns in temporal networks, which include both recurrent and non-recurrent patterns, have been developed and applied to some real-world systems such as social networks \citep{masuda2019detecting, gelardi2021temporal}, human brain networks \citep{lopes2020recurrence}, and human mobility networks \citep{dantsuji2023understanding}.
The evolution of air transport networks has also been investigated, and both abrupt changes and periodic patterns have been quantitatively revealed \citep{sugishita2021recurrence}. 
However, the dynamics of the delay propagation patterns, which also should be affected by the underlying time-varying air transport networks, have not been investigated in the previous studies, and they remain largely unknown.

\subsection{Contributions and outline}
In the present study, we propose a novel framework for analyzing delay propagation patterns, which includes delay causality networks, temporal network analysis, and characterization of the delay propagation patterns. 
We apply the proposed framework to the Japanese domestic air transport network throughout 2019, leading to several key findings. 
The primary novelties in our study can be summarized in the following three points.

First, the most significant novelty of this study lies in investigating the recurrence in the delay propagation patterns by integrating delay causality networks and temporal network analysis.
The existing studies on the delay causality networks \citep{zanin2017network, du2018delay, jia2022delay, guo2022detecting} have utilized traditional methods of static network analysis and analyzed the structure of delay causality networks on each day individually for short periods, around one month. 
These studies do not answer our question on the recurrence of delay propagation patterns.
Our temporal network analysis of the delay causality networks reveals that the delay propagation patterns can be classified into several groups.
We also discover that the identified groups exhibit statistically significant differences in total delay time and average out-degree, with different airports playing central role in spreading delays.
The results also suggest that some delay propagation patterns are more prevalent during specific times of the year, which could be influenced by seasonal and geographical factors in Japan.

Second, we uncover unique characteristics of the delay causality networks by comparing them with randomized networks.
To generate randomized networks, we employ a directed configuration model \citep{newman2001random}, which preserves both in-degree and out-degree distributions, providing a more rigorous analysis than an existing study \citep{du2018delay} that used randomized networks with different in/out-degree distributions.
As a result, delays tend to propagate unidirectionally regardless of the total delays and airlines, which is the opposite conclusion in \cite{du2018delay}. 
In addition, we discover that specific network motifs appear more (or less) frequently than the corresponding randomized counterparts. 
We also find that this characteristic is more pronounced in the groups with more significant delays.
These findings suggest that delays tend to propagate following specific directional patterns, which could significantly contribute to predicting air traffic delays.

Third, we analyze the Japanese domestic air transport network, which has been rarely explored. 
While there are a few studies on air traffic delays in Japan \citep{kato2008analysis, hirata2018flight}, the structure of the delay propagation patterns has never been investigated.
Through this study, we find that Japan's two major full-service carriers, ANA and JAL, have entirely different delay propagation patterns, even though their flight networks are relatively similar throughout the year.
Nevertheless, the aforementioned unidirectionality and network motifs can be observed for both airlines, suggesting that they are universal properties of air traffic delay propagation.

Our paper proceeds as follows. 
In Sec.~\ref{sec:methods}, we describe the proposed framework for analyzing delay propagation patterns. 
In Sec.~\ref{sec:results}, we show the results for the Japanese domestic air transport network in 2019.
In Sec.~\ref{sec:conclusions}, we summarize and discuss our findings and describe future directions.

\section{Methods}
\begin{figure}[ht!]
\centering
\includegraphics[width=\linewidth]{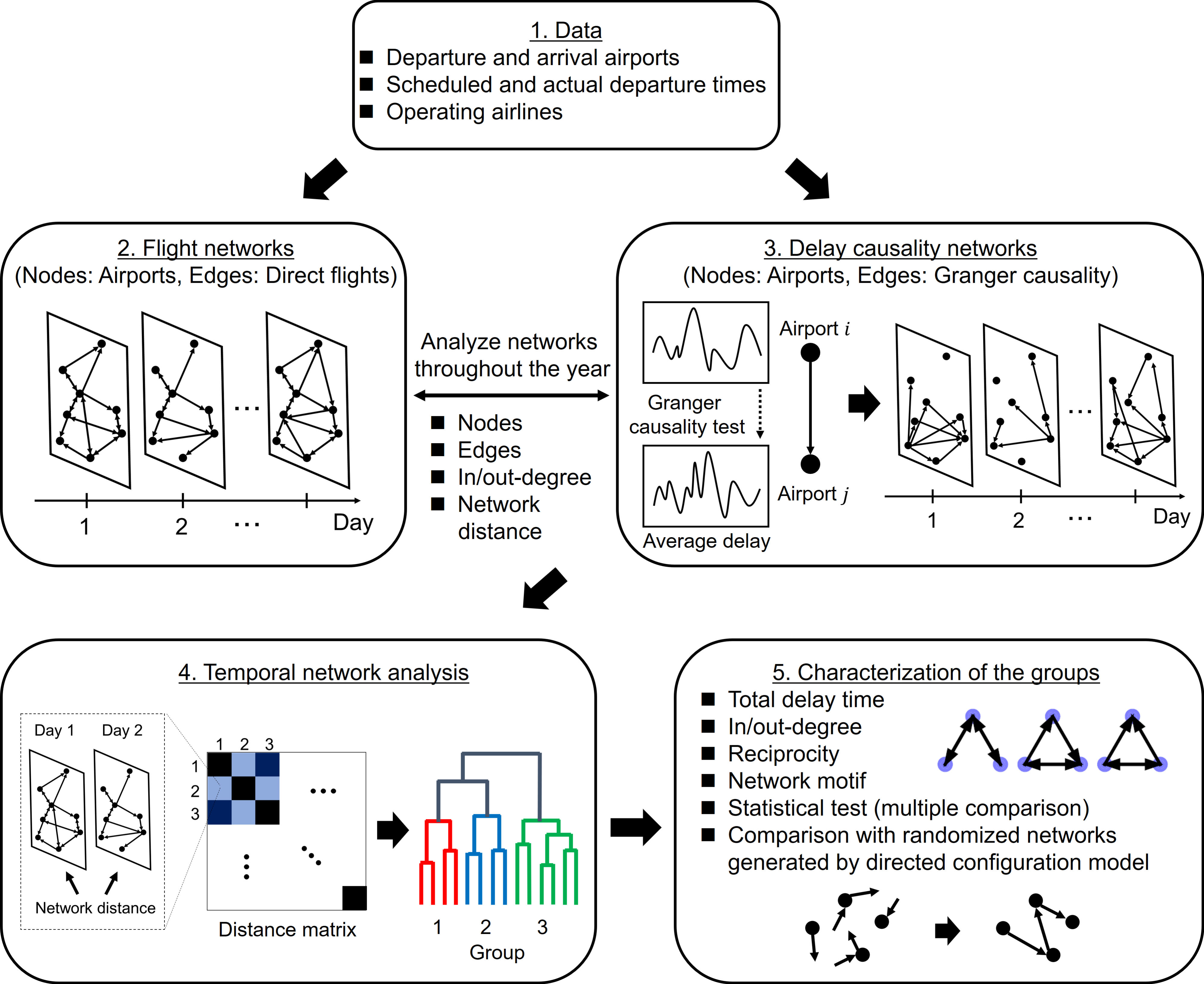}
\caption{Overview of the proposed framework.}
\label{fig:framework}
\end{figure}

We show the overview of the proposed framework in Fig.~\ref{fig:framework}.
The entire framework consists of the construction of both flight networks and delay causality networks, analysis within and between flight networks and delay causality networks, temporal network analysis of the delay causality networks, and characterization of the identified groups of the delay propagation patterns, including comparison with randomized networks generated by directed configuration model.
We describe the details of each step in the following subsections.
\label{sec:methods}

\subsection{Data}

We use data of Japanese domestic flights in 2019 extracted from OAG historical flight data \citep{OAG2022data}. 
The data contains various information about each flight, such as departure and arrival airports, scheduled and actual departure times, and operating airlines, among other details. 
From this data, we construct two kinds of networks: flight networks and delay causality networks. 
We analyze these networks for the entire network, including all airlines and the network of Japan's two largest full-service carriers, ANA and JAL.
Note that any data set other than the OAG historical flight data can be used if it contains the information described above.
For example, for the US air transport networks, one can use an airline on-time performance database, which is open to the public \citep{BTS2023air}.

\subsection{Flight networks}

We construct a daily snapshot of flight networks for Japanese domestic flights in 2019.
In a flight network, nodes are airports, and edges are direct flight connections. 
If there is at least one direct flight from airport $i$ to airport $j$ during the day, we add a directed edge from node $i$ to node $j$.
Therefore, the flight networks are directed and unweighted networks.

\subsection{Delay causality networks}
We construct a daily snapshot of delay causality networks for Japanese domestic flights in 2019.
As described above, we use the Granger causality test to construct the delay causality networks, which has been most commonly used \citep{zanin2017network, du2018delay}.
As described in the following subsections, the delay causality networks are directed and unweighted networks, in which nodes are airports and directed edges are delay causality relationships.

\noindent
\subsubsection*{Delay time series}

We construct a delay time series for each airport on each day.
Specifically, we calculate the average delay on an hourly basis, which is defined by 

\begin{equation}
\label{timeseries}
 d_i(t) = \frac{u_i(t)+c_i(t)h}{s_i(t)}\,, t \in \{6, 7, 8, \cdots, 21\}
\end{equation}
where $d_i(t)$ is the average delay for period $(t, t+1)$ at airport $i$, $u_i(t)$ is the total delay for period $(t, t+1)$ at airport $i$, $c_i(t)$ is the number of canceled flights for period $(t, t+1)$ at airport $i$, $h$ is the equivalent delay time of a canceled flight, and $s_i(t)$ is the total number of flights for period $(t, t+1)$ at airport $i$.
Note that $d_i(t)$ is calculated based on the original departure time of the flight and all time units are denominated in minutes.
We set $h=180$ [min] in the same way as \citet{du2018delay}, which is based on the regulations of the Federal Aviation Administration (FAA), Civil Aviation Administration of China (CAAC), and European Union Aviation Safety Agency (EASA).
Note that we calculate $d_i(t)$ from 6 am to 10 pm because the number of flights for the other periods is very small in our data.

\noindent
\subsubsection*{Granger causality test}

We perform the Granger causality test \citep{granger1969investigating} for a pair of the delay time series for two airports on the same day.
Granger causality is a statistical concept used to determine whether one time series is useful in forecasting another time series. 
Granger causality is based on the idea that if a time series $X$ can predict the future values of another time series $Y$ better than $Y$ alone, then $X$ is said to ``Granger-cause" $Y$.

First, we consider predicting the current value of delay time series $Y_i$ from only the past values of $Y_i$.
We obtain the residual sum of squares, denoted by $SRR_{R}$, with the following restricted regression equation,
\begin{equation}
\label{y_only}
 y_i^T=\sum_{m=1}^{p_{ij}}a^my_i^{T-m}+e^T\,, 
\end{equation}
where $y_i^T$ is the current value of the delay time series $Y_i$, $y_i^{T-m}$ is the past value of the delay time series $Y_i$, $p_{ij}$ is the lag, $e^T$ is the error term, and $a^m$ is a coefficient.

Second, we consider predicting the current value of delay time series $Y_i$ from the past values of both $Y_i$ and $Y_j$.
We obtain the residual sum of squares, denoted by $SRR_{UR}$, with the following unrestricted regression equation,
\begin{equation}
 y_i^T=\sum_{m=1}^{p_{ij}}a^my_i^{T-m}+\sum_{m=1}^{p_{ij}}b^my_j^{T-m}+e^T\,, 
\end{equation}
where $y_j^{T-m}$ is the past value of the delay time series $Y_j$ and $b^m$ is a coefficient.
The null hypothesis that the delay time series $Y_j$ does not cause the delay time series $Y_i$ is represented by 
\begin{equation}
 b^1=b^2=\cdots=b^{p_{ij}}=0\,. 
\end{equation}

Finally, we test the null hypothesis with the following F-statistics,
\begin{equation}
F=\frac{(RSS_R-RSS_{UR})/p_{ij}}{RSS_{UR}/(w-p_{ij})}\,,
\end{equation}
where $w$ is the sample size of each time series.
The null hypothesis is rejected when the p-value is less than the significance level ($5\%$ in this study).
This implies that the Granger causality exits from the delay time series $Y_j$ to the delay time series $Y_i$.

\noindent
\subsubsection*{Construction of delay causality networks}

We perform the Granger causality test for all the possible pairs of the delay time series for different airports on the same day, $N*(N-1)$ times, where $N$ is the number of airports.
If we observe the existence of the Granger causality from airport $i$ to airport $j$, we add a directed edge from node $i$ to node $j$.
The resulting network is a delay causality network.
We construct the delay causality network for each day in 2019.

\subsection{Temporal network analysis}

\noindent
\subsubsection*{Network distance and distance matrix}

We explore the evolution of the delay propagation patterns by measuring dissimilarity in the structure of the delay causality networks on different days.
To measure the dissimilarity between snapshot networks on day $i$ and day $j$, we define the network distance between two networks $G_i$ and $G_j$ as
\begin{equation}
L(G_i,G_j)= 1-\frac{M(G_i\cap G_j)}{\sqrt{M(G_i)M(G_j)}}\,,
\label{eq:normalized_network_distance}
\end{equation}
where $M(G_i)$ and $M(G_j)$ are the numbers of directed edges in $G_i$ and $G_j$ respectively, and $M(G_i\cap G_j)$ is the number of directed edges that $G_i$ and $G_j$ have in common \citep{sugishita2021recurrence}. 
Network distance $L$ ranges between $0$ and $1$. 
When we measure the network distance between all pairs of days in 2019, the resulting distance matrix is a $365 \times 365$ symmetric matrix of which the $(i, j)$th entry is given by $L(G_i, G_j)$, where $G_i$ is the network on day $i$.

\noindent
\subsubsection*{Classification of delay propagation patterns}

\label{sec:classification}
We classify the delay propagation patterns by applying a framework for detecting states in temporal networks \citep{masuda2019detecting} to the constructed delay causality networks.
Here, we have a sequence of 365 snapshot delay causality networks. 
To assign a state to each of the 365 snapshot networks, we apply a hierarchical clustering algorithm to the $365 \times 365$ distance matrix. 
We used ``linkage'' and ``fcluster'' functions in the ``scipy'' module in Python. 
We used ``ward'' in the ``linkage'' function to define the distance between clusters. 
The hierarchical clustering divides the delay causality networks into $C$ discrete states, where the number of states, $C$, ranges between $1$ and $365$. 
We adopt the value of $C\ (2 \le C \le 365$ that maximizes the Dunn's index, $D$ \citep{dunn1974fuzzy}, defined by
\begin{equation}
D=\frac{\min_{1 \leq c \neq c' \leq C} \min_{G_i \in c{\rm th\ state}, G_j \in c'{\rm th\ state}} L(G_i,G_j)}{\max_{1 \leq c'' \leq C} \max_{G_{i'}, G_{j'} \in c''{\rm th\ state}} L(G_{i'},G_{j'})}\,.
\label{eq:dunn}
\end{equation}
The numerator on the right-hand side of Eq.~\eqref{eq:dunn} represents the smallest distance between two states among all the pairs of states. The denominator represents the largest diameter of the state among all the states.

\subsection{Characterization of the groups}

We characterize the identified groups regarding the total delay, average out-degree, central airports of spreading delays, reciprocity, and network motifs.
For the reciprocity and network motifs, we compare the delay causality networks with corresponding randomized networks generated by the directed configuration model to reveal unique characteristics of the delay causality networks that deviate from the randomized networks.

\noindent
\subsubsection*{In-degree and out-degree}

In-degree and out-degree are the number of edges pointing inward to and outward from the nodes, respectively \citep{newman2018networks}.
An airport with a large in-degree in the delay causality network indicates that delays are being spread out from many other airports to that airport.
On the other hand, an airport with a large out-degree in the delay causality network indicates that such an airport spreads delay to many other airports.

Mathematical definitions of the in-degree and out-degree are as follows.
Let $A_{ij}$ be the adjacency matrix of a directed network. 
The in-degree of node $i$, denoted by $k_{i}^{\rm in}$, is given by
\begin{equation}
    k_{i}^{\rm in} = \sum_{j=1}^{N} A_{ji}\,.
\end{equation}
The average in-degree over all nodes is given by 
\begin{equation}
    \langle k^{\rm in} \rangle = \frac{1}{N}\sum_{i=1}^{N} k_{i}^{\rm in}\,.
\end{equation}
Similarly, the out-degree of node $i$, denoted by $k_{i}^{\rm out}$, is given by
\begin{equation}
    k_{i}^{\rm out} = \sum_{j=1}^{N} A_{ij}\,.
\end{equation}
The average out-degree over all nodes is given by 
\begin{equation}
    \langle k^{\rm out} \rangle = \frac{1}{N}\sum_{i=1}^{N} k_{i}^{\rm out}\,.
\end{equation}

\noindent
\subsubsection*{Reciprocity}

If there is an edge from node $i$ to node $j$ and an edge from node $j$ to node $i$, we say such an edge is reciprocated \citep{newman2018networks}.
The reciprocity, denoted by $R$, is defined as the fraction of edges that are reciprocated:
\begin{equation}
    R=\frac{1}{M}\sum_{ij} A_{ij}A_{ji}\,,
\end{equation}
where $M$ is the number of edges. 
Note that the product of adjacency matrix elements, $A_{ij}A_{ji}$, is 1 if and only if there is an edge from node $i$ to node $j$ and an edge from node $j$ to node $i$, and is zero otherwise.
The reciprocity can take values between 0 and 1; larger values indicate more bidirectional networks. 
In this study, we compare the reciprocity between the delay causality network and the corresponding randomized networks generated by the directed configuration model to extract unique characteristics of the delay causality networks.

\noindent
\subsubsection*{Network motif}

\begin{figure}[b!]
\centering
\includegraphics[width=0.9\linewidth]{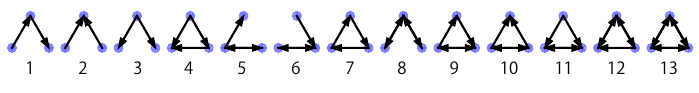}
\caption{All 13 types of three-node subgraphs.}
\label{fig:three-node_subgraph}
\end{figure}

Network motifs are recurrent and statistically significant subgraphs of a large network \citep{milo2002network}.
We scan all possible three-node subgraphs and record the number of occurrences of each subgraph.
As shown in Fig.~\ref{fig:three-node_subgraph}, there are 13 types of the three-node subgraphs.
In the same way as the reciprocity, we compare the number of occurrences of each subgraph with that of corresponding randomized networks generated by the directed configuration model.
Note that while the subgraphs that appear more frequently than expected in the randomized networks are usually focused on, we also focus on the subgraphs that occur less often than expected in the randomized counterparts because they are also valuable for understanding the dynamics of delay propagation.

\noindent
\subsubsection*{Statistical analysis for multiple comparison}

Here, we describe the statistical analysis to compare the measurements (total delay time and average out-degree in this study) among the identified delay propagation patterns. 
First, we use the Kruskal-Wallis, a nonparametric statistical test, to assess the differences among three or more groups \citep{kruskal1952use}.
We used the ``Kruskal'' function in a Python package ``SciPy Stats''.
Second, as a post-hoc test, we use the Steel-Dwass test \citep{steel1960rank} to compare all pairs of the groups.
We used the ``posthoc\_dscf'' function in a Python package ``scikit-posthocs''.

\noindent
\subsubsection*{Directed configuration model}

\label{sec:directed_configuration_model}
We examine the unique characteristics of delay causality networks by comparing them with randomized networks. 
In this study, we use a directed configuration model \citep{newman2001random} to generate randomized networks, which rewires edges randomly while preserving both the in-degree and out-degree of each node. 
Therefore, we can compare the network measurements (reciprocity and network motifs
in this study) between the original delay causality networks and the corresponding randomized networks, which have the same in-degree and out-degree distributions.

The steps for generating a randomized network for a delay causality network are as follows.
First, we select two directed edges $(a, b)$ and $(c, d)$ uniformly at random in the original delay causality network.
We then rewire them to create new directed edges $(a, d)$ and $(c, b)$.
This rewiring preserves both the in-degree and out-degree of each node.
Note that if the rewiring generates self-loop(s) or multiple edge(s), we discard the attempt and try again with a new random selection of edges.
We repeat this process 1,000 times to generate a randomized network.

We generate 1,000 randomized networks and compare the network measurements between the original delay causality network and the 1,000 randomized networks. 
Specifically, we calculate the z-score of the value of the original delay causality network to the values of the 1,000 randomized networks.   
The z-score, denoted by $z$, is given by \begin{equation}
    z=\frac{x_{\rm DCN}-\mu_{\rm rand}}{\sigma_{\rm rand}}\,,
\end{equation}
where $x_{\rm DCN}$ is the value of the original delay causality network, and $\mu_{\rm rand}$ and $\sigma_{\rm rand}$ are the mean and the standard deviation of the 1,000 randomized networks, respectively.

\section{Results}
\label{sec:results}

\subsection{Flight and delay causality networks throughout 2019}
\begin{figure}[b!]
\centering
\includegraphics[width=\linewidth]{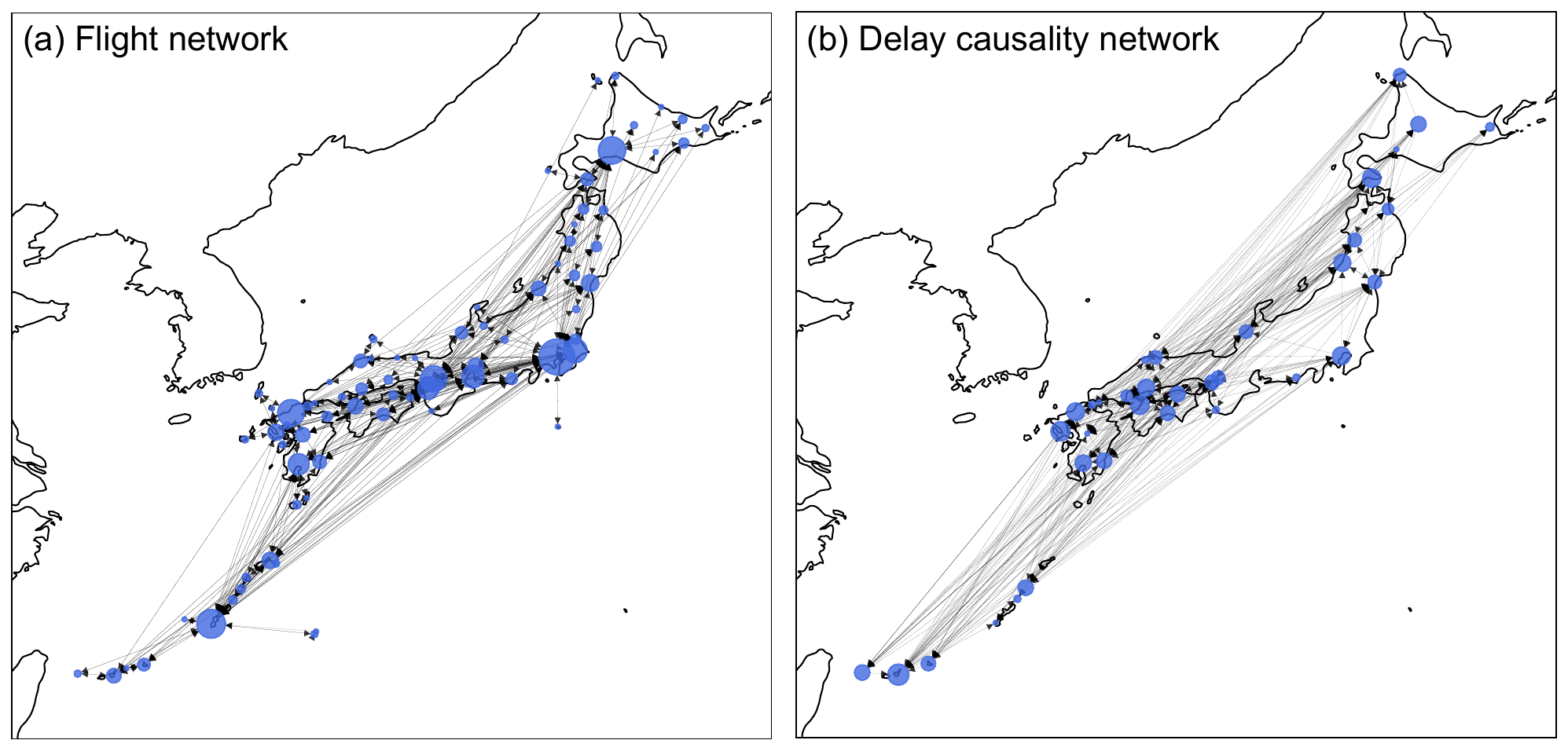}
\caption{Flight and delay causality networks for the Japanese domestic flights on January 1st, 2019. (a) Flight network. Nodes are airports, and edges are direct flight connections. (b) Delay causality network. Nodes are airports, and edges are Granger causality. In both networks, the size of nodes is proportional to their out-degree.}
\label{fig:network_january_1}
\end{figure}

We show the flight and delay causality networks for January 1st, 2019, as an example in Fig.~\ref{fig:network_january_1}. 
Recall that in both networks, the nodes represent airports, but the edges in the flight network are direct flights, and the edges in the delay causality network are Granger causality.
On this day, the number of nodes in the flight network is 78, while that of the delay causality network is 37, which indicates that delay propagation occurs at about half of the airports.

\begin{figure}[ht!]
\centering
\includegraphics[width=0.9\linewidth]{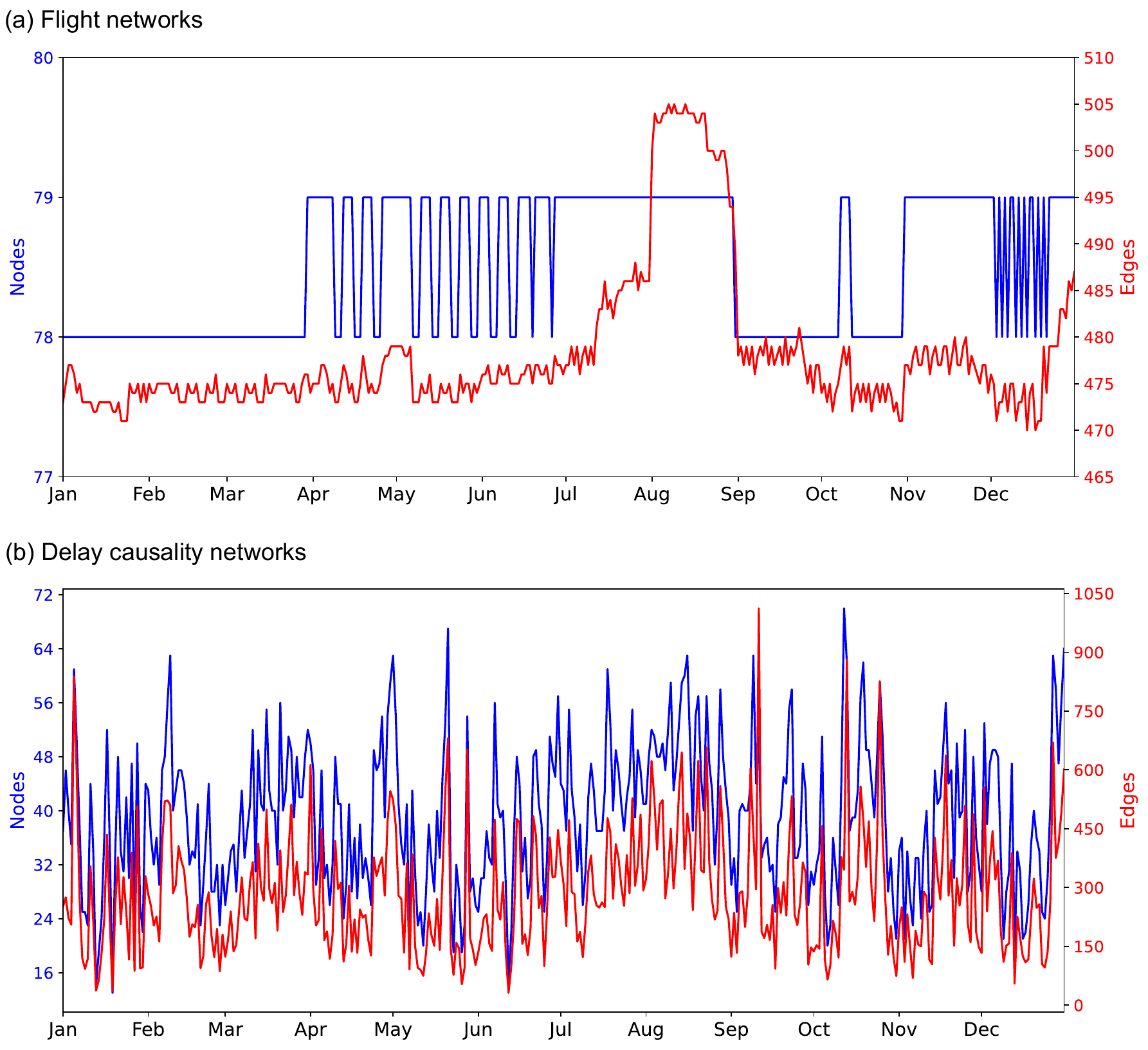}
\caption{Evolution of nodes and edges in the (a) flight networks and (b) delay causality networks for the Japanese domestic flights in 2019.}
\label{fig:entire_network}
\end{figure}

Figure~\ref{fig:entire_network} shows the evolution of nodes and edges for the flight and delay causality networks in 2019. 
In the flight networks, the number of nodes is very stable throughout the year, taking values of 78 or 79. 
On the other hand, the number of edges shows both small weekly and relatively large seasonal variations. 
Seasonal variations include increases in August and at the end of December. 
Passenger demand is high during these periods due to major holidays.

In contrast, the number of nodes and edges fluctuates significantly daily in the delay causality networks. 
No obvious periodic patterns can be observed from the visual inspection. 
It is also clear that the number of nodes and edges in the delay causality network are not exceptionally high at the end of August and December when the number of edges in the flight networks is high.
Therefore, the number of nodes and edges in the delay causality networks alone cannot determine whether specific patterns exist.

We show the Pearson correlation coefficient between in-degree and out-degree in the flight and delay causality networks for all airlines, ANA, and JAL in Table \ref{tab:correlation}.
We first notice that there is a moderate positive correlation between the in-degree and out-degree of the delay causality networks, i.e., $\langle k_{\rm DCN}^{\rm in}\rangle$ and $\langle k_{\rm DCN}^{\rm out}\rangle$.
This suggests that airports that receive delay propagation from many airports tend to spread out delays to many airports.
We also notice that the value for all airlines is more significant than those for ANA and JAL.
This implies that smaller airlines other than ANA and JAL, including low-cost carriers, may exhibit a stronger correlation in delay propagation. 

\begin{table}[t!]
\centering
\caption{Correlation coefficient between in-degree and out-degree in the flight and delay causality networks on the same day for all airlines, ANA, and JAL. 
The values are $\rm mean \pm \rm standard\, deviation$ over 365 days in 2019.
FN and DCN represent flight networks and delay causality networks, respectively.
}
\label{tab:correlation}
\begin{tabular}{ccccc}
\cline{1-4}
                    & All airlines    & ANA            & JAL            &  \\ \cline{1-4}
$\langle k_{\rm DCN}^{\rm in} \rangle$ vs $\langle k_{\rm DCN}^{\rm out} \rangle$ & $0.659 \pm 0.0983$ & $0.594 \pm 0.156$ & $0.532 \pm 0.180$ &  \\
$\langle k_{\rm FN}^{\rm in} \rangle$ vs $\langle k_{\rm DCN}^{\rm in} \rangle$   & $0.289 \pm 0.119$  & $0.334 \pm 0.170$ & $0.230 \pm 0.202$ &  \\
$\langle k_{\rm FN}^{\rm out} \rangle$ vs $\langle k_{\rm DCN}^{\rm out} \rangle$ & $0.269 \pm 0.123$  & $0.306 \pm 0.163$ & $0.233 \pm 0.201$ &  \\ \cline{1-4}
\end{tabular}
\end{table}
On the other hand, we found weak positive correlations between the in-degree of the flight networks, $\langle k_{\rm FN}^{\rm in}\rangle$, and that of the delay causality networks, $\langle k_{\rm DCN}^{\rm in}\rangle$. 
We obtain similar results for the out-degree, i.e., $\langle k_{\rm FN}^{\rm out}\rangle$ and $\langle k_{\rm DCN}^{\rm out}\rangle$. 
These results indicate that hubs in the flight networks are not necessarily central airports in terms of delay propagation.

\begin{figure}[t!]
\centering
\includegraphics[width=0.8\linewidth]{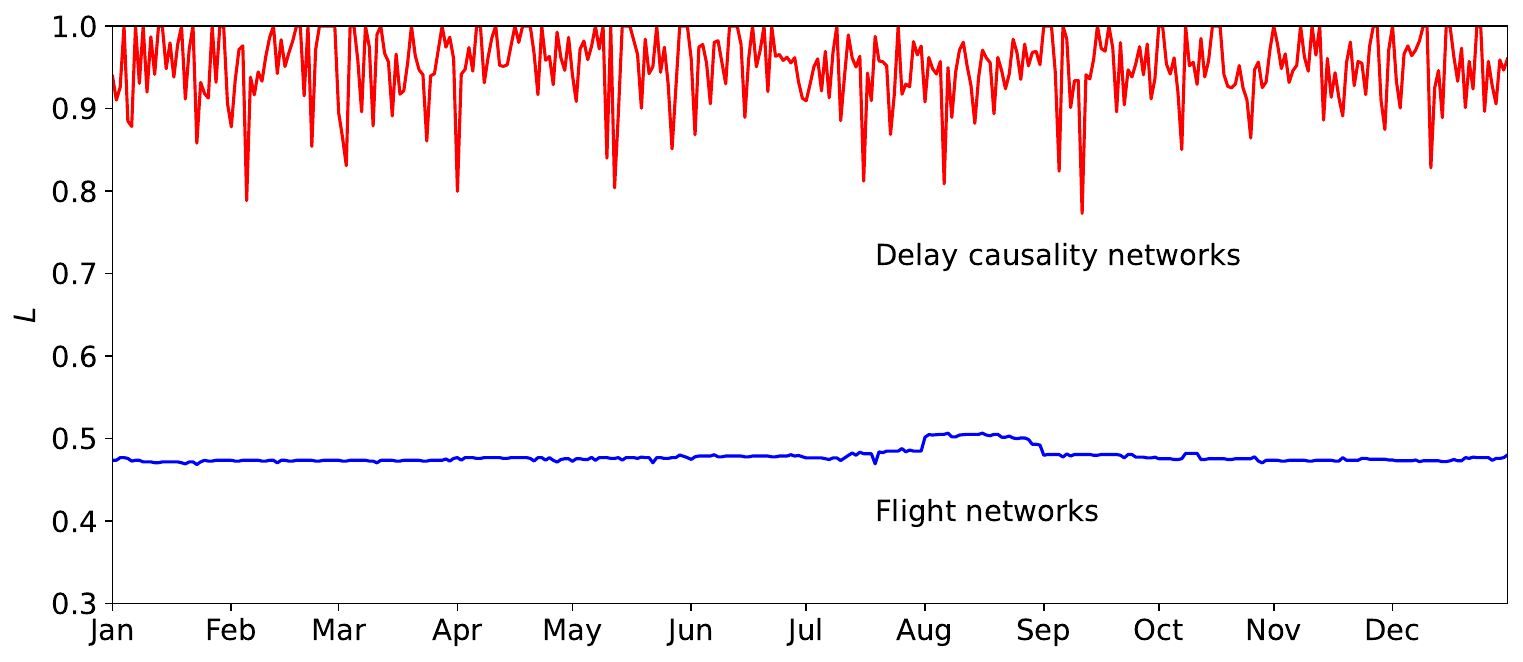}
\caption{Network distance of the flight and delay causality networks between ANA and JAL in 2019.}
\label{fig:ANA_JAL}
\end{figure}

We compare the structure of the flight and delay causality networks between ANA and JAL.
Figure~\ref{fig:ANA_JAL} shows the network distance $L$ of the flight and delay causality networks between ANA and JAL for each day in 2019.     
First, the network distance of the flight networks between ANA and JAL is stable throughout the year, taking values of about 0.5. 
In contrast, interestingly, the network distance of the delay causality networks takes much larger values, around 0.9, with more significant fluctuations. 
These results suggest that even though the flight networks of ANA and JAL are relatively similar, their delay propagation patterns are entirely different.

\subsection{Classification of the delay propagation patterns}

\begin{figure}[ht!]
\centering
\includegraphics[width=0.75\linewidth]{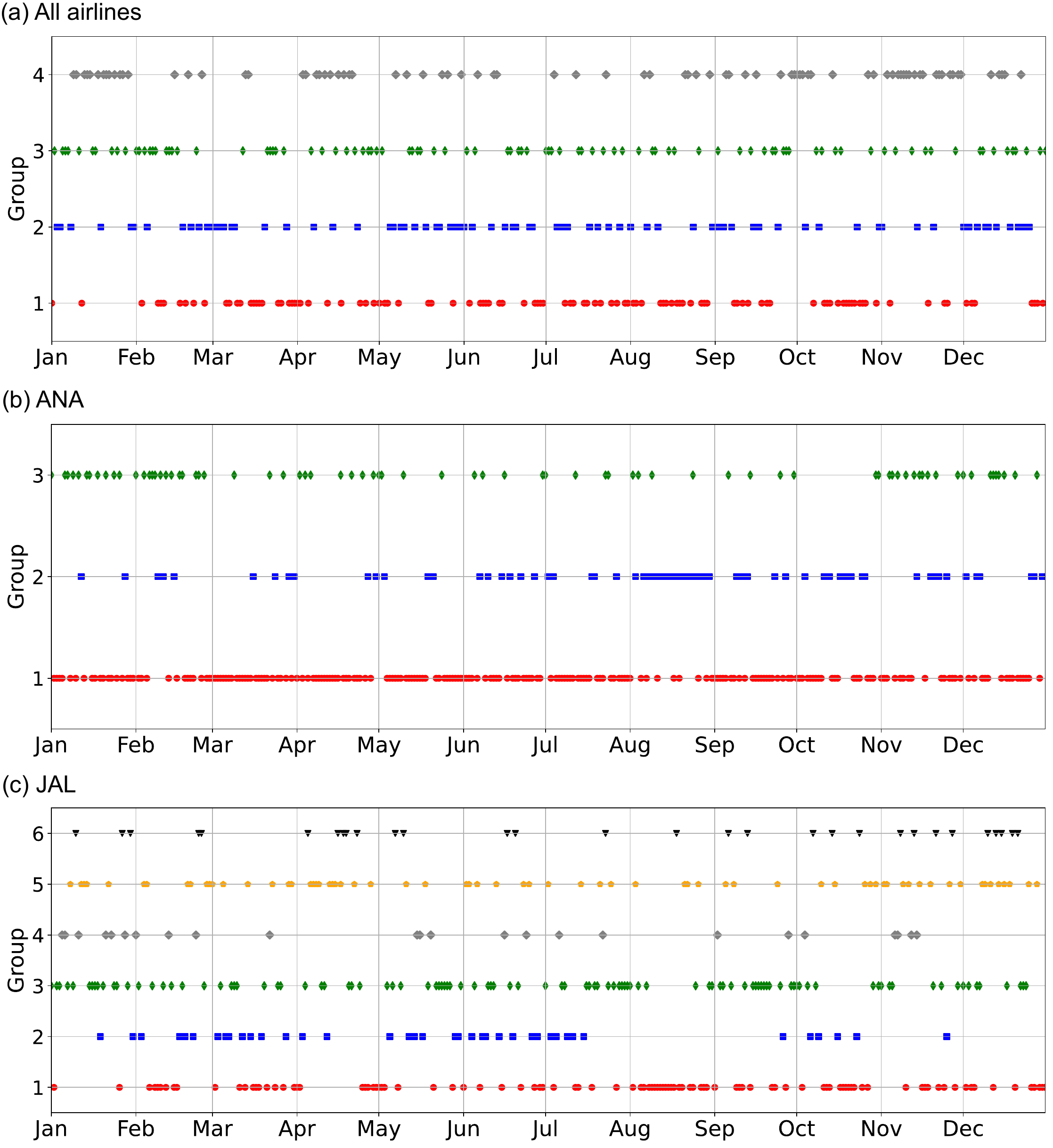}
\caption{The appearances of the identified groups over days for (a) all airlines, (b) ANA, and (c) JAL.
}
\label{fig:days}
\end{figure}
\add{In the previous section, we analyzed the structure of the flight and delay causality networks throughout 2019. 
Here, we classify the structure of the delay causality networks by temporal network analysis. 
The results suggest that the delay causality networks can be classified  into four, three, and six groups for all airlines, ANA, and JAL, respectively (see the distance matirices and dendrogram based on the hierarchical clustering in Figs.~\ref{fig:distance_matrix} and \ref{fig:dendrogram}, respectively).}
Figure~\ref{fig:days} illustrates the appearances of each group over days.
Notably, in the results for ANA (Fig.~\ref{fig:days}(b)), group 2 includes many days in June and August.
In contrast, group 3 prominently features January, February, November, and December days. 
Therefore, in broad terms, group 2 represents delay propagation patterns frequently occurring in the summer, while group 3 embodies delay propagation patterns commonly seen in the winter. 
Turning our attention to the results for JAL (Fig.~\ref{fig:days}(c)), although there is not a clear seasonality as seen in ANA, it is noticeable that group 1 includes most of the days in August. 
Observing results from all airlines (Fig.~\ref{fig:days}(a)), there is no apparent seasonality. 
These findings suggest that some airlines have delay propagation patterns that are more prevalent during specific times of the year.

\subsection{Characterization of the groups}
We further characterize the groups identified by the temporal network analysis for all airlines, ANA, and JAL in terms of the total delay, average out-degree, central airports to spread delays, reciprocity, and network motifs.

\noindent
\subsubsection*{Total delay time}

\begin{figure}[t!]
\centering
\includegraphics[width=\linewidth]{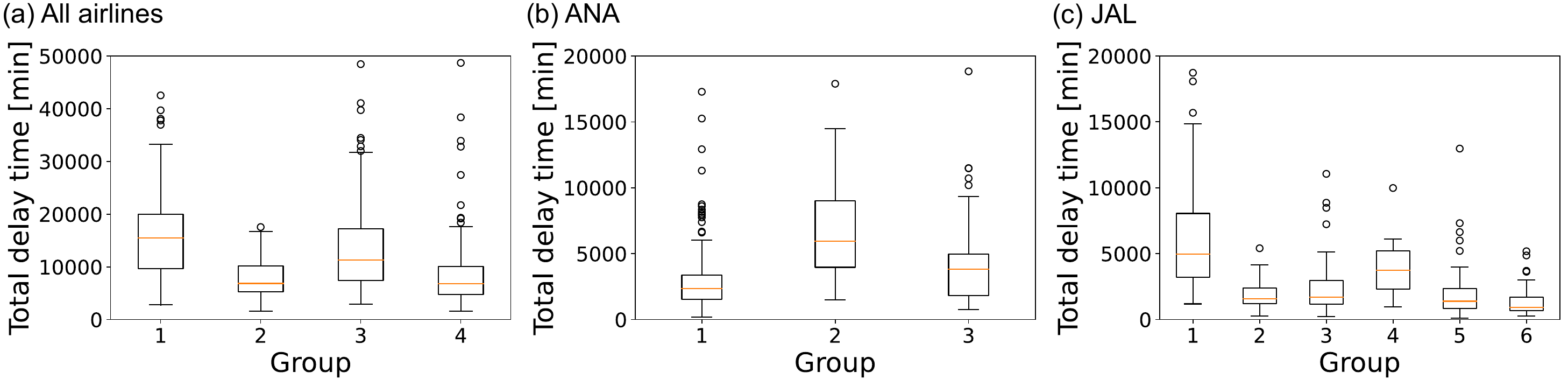}
\caption{Comparison of the total delay time among the identified groups for (a) all airlines, (b) ANA, (c) JAL. 
We use box plots to show five-number summaries of the distributions; these quantities are the first quartile ($Q_1$), the median, the third quartile ($Q_3$), the minimum without outliers ($Q_1-1.5 \, \times \, {\rm IQR}$), and the maximum without outliers ($Q_3+1.5 \, \times \, {\rm IQR}$), where ${\rm IQR}=Q_3-Q_1$. Open circles indicate outliers.
}
\label{fig:total_delay_time}
\end{figure}

We compare the total delay time, defined as the sum of the delay on a day, among the identified groups.
First, we carry out the Kruskal–Wallis test, which reveals a significant difference among the distributions of the total delay time for all the three cases, i.e., all airlines, ANA, and JAL, shown in Fig.~\ref{fig:total_delay_time} ($p<0.001$).
Second, as a post-hoc test, we run the Steel-Dwass test to compare all the pairs of groups. 
As a result, we found statistically significant differences ($p<0.05$) for all the pairs except for the following pairs: 2-4 of all airlines; 1-4, 2-3, 2-5, 2-6, 3-5, and 5-6 of JAL.  
Interestingly, even though the classification was based solely on the structure of the delay causality networks, we found statistically significant differences in terms of the total delay time, which is not directly related to the network structure.

\begin{figure}[t!]
\centering
\includegraphics[width=\linewidth]{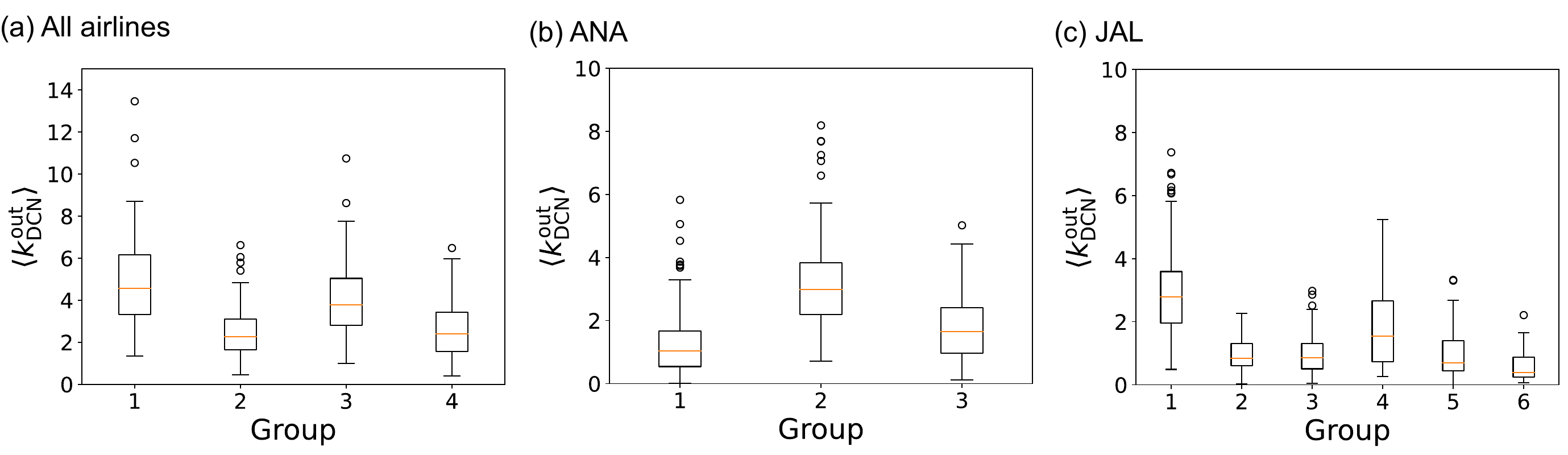}
\caption{Comparison of the average out-degree among the identified groups for (a) all airlines, (b) ANA, (c) JAL.
}
\label{fig:ave_out_degree}
\end{figure}

\noindent
\subsubsection*{Average out-degree and central airports of spreading delays}

We show the distribution of the average out-degree in Fig.~\ref{fig:ave_out_degree}.
In the same way as the total delay time, we first carry out the Kruskal–Wallis test, which reveals a significant difference among the distributions of the average out-degree  for all three cases, i.e., all airlines, ANA, and JAL ($p<0.001$).
Second, as a post-hoc test, we run the Steel-Dwass test to compare all the pairs of groups. 
As a result, we found statistically significant differences ($p<0.05$) for all the pairs except for the following pairs: 2-4 of all airlines: 2-3, 2-4, 2-5, 3-5, and 5-6 of JAL.  
We notice that these results are similar to the total delay time results. 
These results suggest that the delay causality networks for the days with considerable total delays tend to be dense.

\begin{figure}[t!]
\centering
\includegraphics[width=0.85\linewidth]{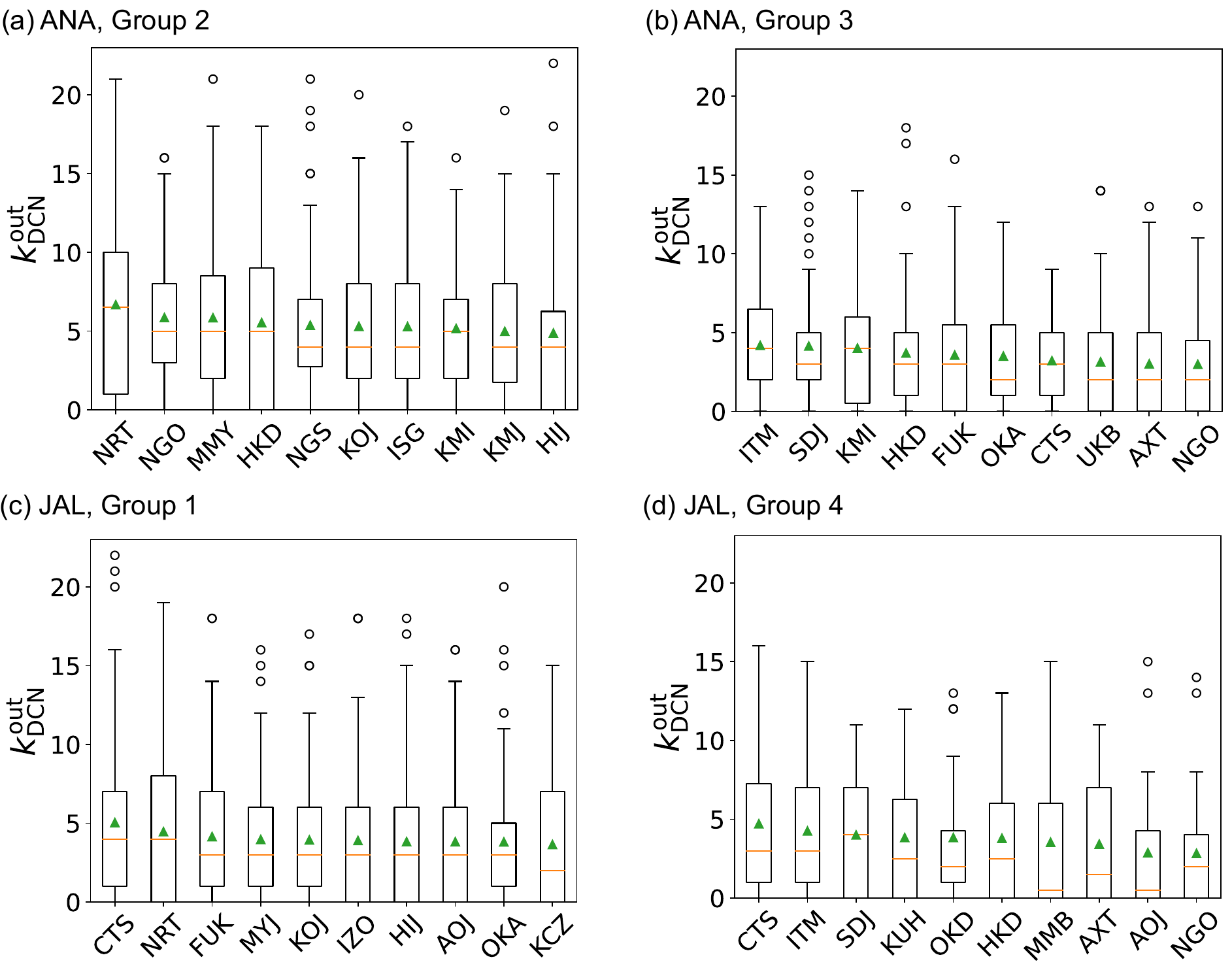}
\caption{Airports with the ten largest out-degree for the two largest delay groups of ANA and JAL. (a) ANA, group 2. (b) ANA, group 3.  (c) JAL, group 1. (d) JAL, group 4.
}
\label{fig:top10_outdegree}
\end{figure}

We examine which airports play the central roles of delay propagation in the two largest delay groups for ANA and JAL. 
Figure~\ref{fig:top10_outdegree} shows the top 10 airports with the largest out-degree.
First, we compare the airports between group 2 and group 3 in ANA (Figs.~\ref{fig:top10_outdegree}(a) and \ref{fig:top10_outdegree}(b)).
We notice that the top 10 airports are primarily different between them.
We found that most of the top 10 airports in group 2 are airports on the islands of Okinawa prefecture, i.e., Miyako (MMY) and New Ishigaki (ISG), and airports in Kyushu island, i.e., Nagasaki (NGS), Kagoshima (KOJ), Miyazaki (KMI), and Kumamoto (KMJ).
Because the Japanese archipelago extends from the subarctic north to the subtropical south, it spans various climate zones. 
Therefore, weather conditions vary greatly depending on the region.
Strong typhoons frequently hit Okinawa and Kyushu islands compared to the rest of the country.
It is conceivable that these typhoons primarily affect the delays.
These typhoons mainly occur in the summer, which is consistent with the frequent appearances of group 2 during the summer (Fig.~\ref{fig:days}(b)).
The impact of climate and typhoons on delays in the Kyushu and Okinawa regions has been pointed out \citep{MLIT2024delay}, which also supports our results.
On the other hand, group 3 tends to include large airports such as Osaka  (ITM), Fukuoka (FUK), Naha (OKA), and New Chitose (CTS). 
Recalling that group 3 frequently appears during the winter (Fig.~\ref{fig:days}(b)) and has the second-largest delays (Fig.~\ref{fig:total_delay_time}(b)), it can be said that delays in the winter tend to propagate across the country, with significant airports serving as the epicenters of the delay spreading.

Next, we compare groups 1 and 4 of JAL (Figs.~\ref{fig:top10_outdegree}(c) and \ref{fig:top10_outdegree}(d)). 
Both groups have CTS at the top of the list, but the rest of the airports differ significantly between them. 
The top 10 airports in group 1 are widely dispersed across the country.
In contrast, for group 4, all the airports except Osaka (ITM) and Chubu Centrair International Airport (NGO) are located in the Tohoku or Hokkaido regions.
Group 4, alongside group 1, has the most significant delays and tends to appear frequently in January (Fig.~\ref{fig:days}(c)). 
Tohoku and Hokkaido are significantly affected by heavy snowfalls in the winter, which are likely to contribute substantially to these delays \citep{MATAYOSHI2012KJ00008275014, MLIT2024delay}.
It should be noted that Haneda Airport, Japan's largest hub airport, is not ranked in the top 10 for both ANA and JAL.

\noindent
\subsubsection*{Directionality in the delay propagation}

\begin{figure}[b!]
\centering
\includegraphics[width=\linewidth]{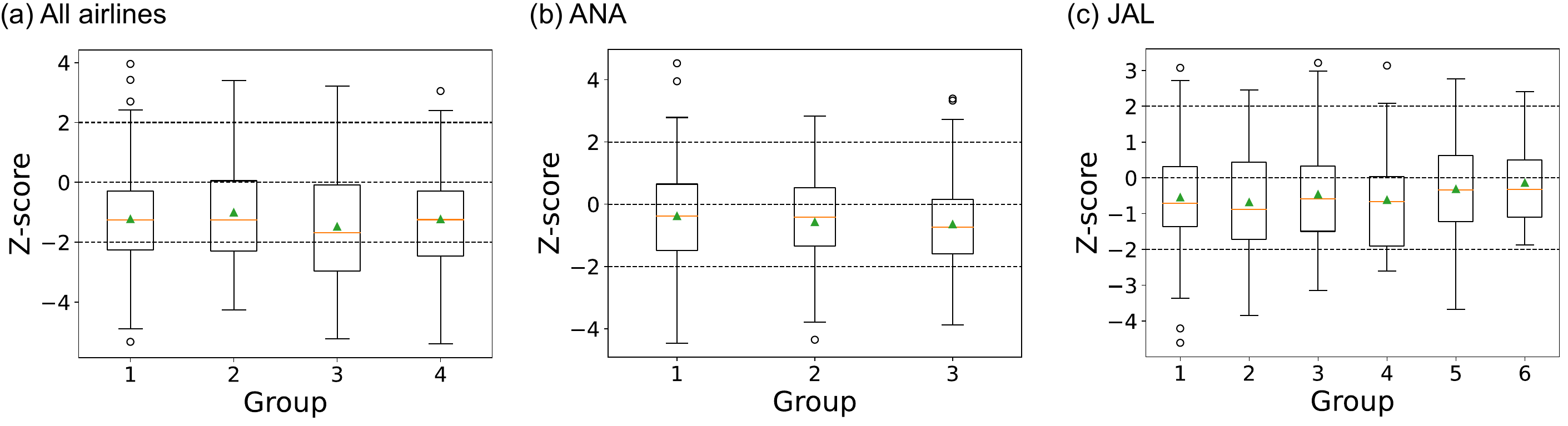}
\caption{Z-score of the reciprocity among the identified groups for (a) all airlines, (b) ANA, (c) JAL.
}
\label{fig:reciprocity}
\end{figure}

Figure~\ref{fig:reciprocity} shows the distribution of the z-score of the reciprocity for each group.
We found that the z-scores tend to be negative across all the groups.
This result suggests that the delay propagation tends to be unidirectional regardless of the amount of delays and the airlines.

\begin{figure}[t!]
\centering
\includegraphics[width=0.75\linewidth]{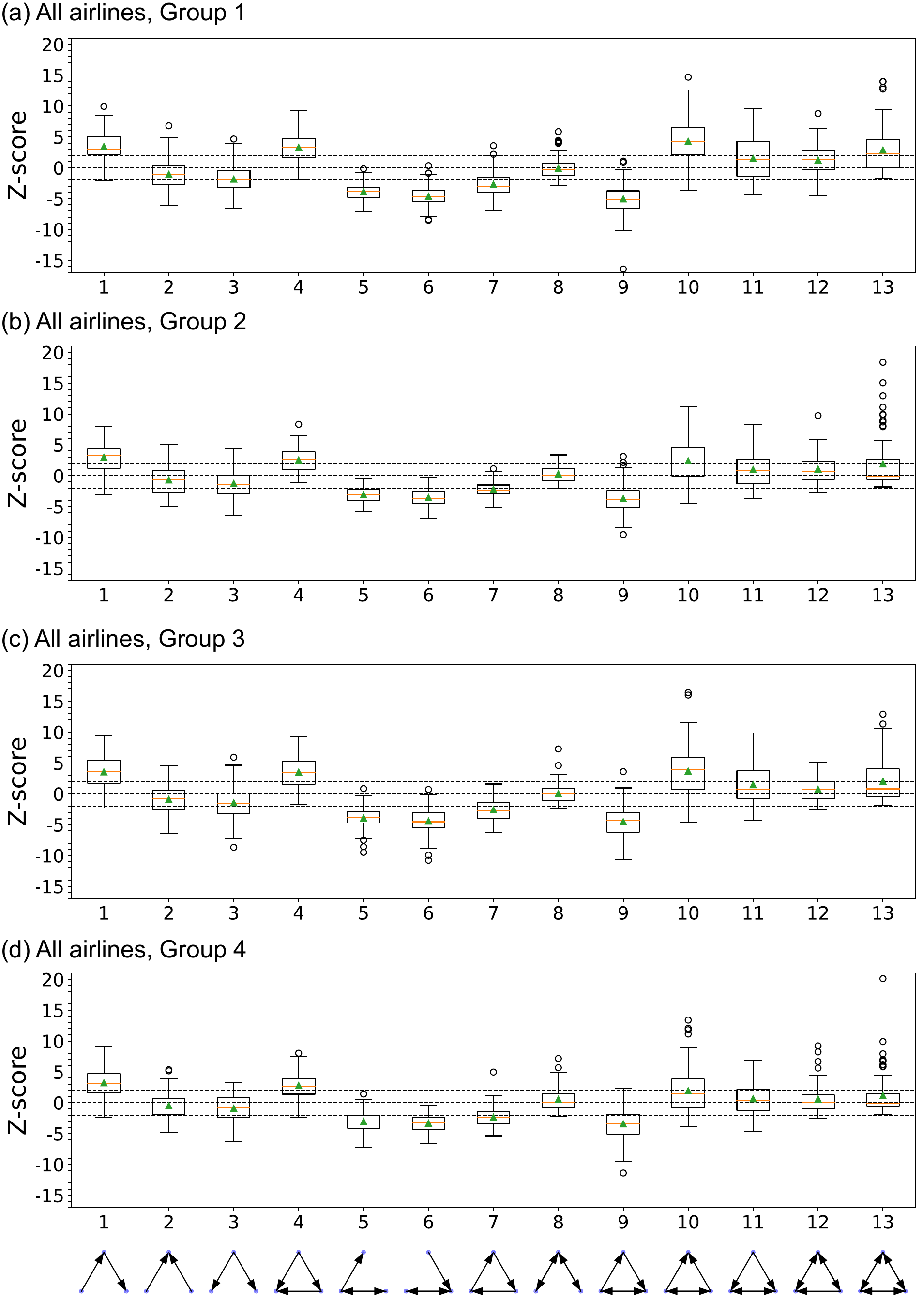}
\caption{Z-score of the network motifs for each group of all airlines.
}
\label{fig:motif}
\end{figure}

In Fig.~\ref{fig:motif}, we show the distribution of the z-score of network motifs for each group for all airlines.
We first notice that all the groups exhibit similar distributions. 
Specifically, the network motifs 1, 4, and 10 appear more frequently in the delay causality networks than the corresponding randomized networks.
On the other hand, the network motifs 5, 6, 7, and 9 tend to appear less frequently in the delay causality networks than the randomized counterparts.
Recalling that groups 1 and 3 for all airlines have more significant total delay times (see Fig.~\ref{fig:total_delay_time}(a)), we further notice that these characteristics are especially pronounced in the groups with more significant delays.
We obtain similar results for ANA and JAL (see Figs.~\ref{fig:motif_ANA} and \ref{fig:motif_JAL}).

The network motif 1, often called \textit{three chain}, appears more frequently than expected in the corresponding randomized networks. 
This network motif represents a pattern where a delay at one airport $a$ will likely cause a delay at a second airport $b$, which then cause a delay at a third airport $c$ ($a \rightarrow b \rightarrow c$).
This result suggests that the delays often propagate linearly from one airport to the next in an unidirectional way.
Like network motif 1, network motif 4, often called \textit{feed-forward loop}, also tends to appear more frequently than expected in the randomized networks.
This network motif represents a pattern where a delay from an airport $a$ influences other airports $b$ and $c$ directly ($a\rightarrow c$) and indirectly ($a\rightarrow b \rightarrow c$).
Thus, similarly to network motif 1, this result also suggests that delays tend to propagate in a unidirectional way. 
In contrast to the network motif 1 and 4, the network motif 7, often called \textit{three-node feedback loop}, tends to appear less frequently than expected in the randomized networks.
This result suggests that delays are less likely to propagate cyclic ($a\rightarrow b \rightarrow c \rightarrow a$), further supporting the unidirectional nature observed in the network motifs 1 and 4.

The network motif 10, often called \textit{uplinked mutual dyad}, appears more frequently than expected in the corresponding randomized counterparts.
This motif represents a pattern where two airports $a$ and $b$ are mutually connected ($a \leftrightarrow b$), and each of them is further connected to airport $c$ ($a \rightarrow c$ and $b \rightarrow c$).
This result is even more intriguing when we consider the results for the network motifs 5, 6, and 9. 
In contrast to the network motif 10, these three motifs appear less frequently than in the randomized networks. 
A common feature of these motifs is the presence of a bidirectional relationship. 
The result for the network motif 5 implies that when there is a bidirectional edge, it is less likely that a delay propagates from just one airport to a third airport. 
In addition, the result for the network motif 6 suggests that when there is a bidirectional edge, it is less likely that a delay propagates from a third airport to just one of the other two airports.
Moreover, the network motif 9, which combines these two motifs, also appears less frequently.
All these results imply that if delays propagate in both directions between two airports, they are likely to spread further to another airport from both airports.

\section{Conclusion}
\label{sec:conclusions}

We have proposed a novel framework for analyzing delay propagation patterns and applied it to the Japanese domestic air transport network in 2019.
Our studies led to several key findings and new insights valuable for understanding air traffic delay propagation.

We first observed that delay causality networks fluctuate significantly daily, whereas the flight networks remain relatively stable.
In addition, the delay causality networks do not show any apparent periodicity, even though the flight networks show weekly periodic variations in the number of edges.
However, the temporal network analysis of the delay causality networks revealed a specific recurrence in the delay propagation patterns: we discovered that the delay causality networks can be classified into several groups. 
The identified groups demonstrated statistically significant differences in the total delay time.
This result is interesting because our classification is solely based on the structure of the delay causality networks. 
Additionally, the groups also have statistically significant differences in the average out-degree. 
We also found a similar tendency between the distributions of the total delay time and the average out-degree, suggesting that the delay causality networks tend to be dense on days with significant delays.

We also revealed unique characteristics in the delay causality networks that deviated from the corresponding randomized networks. 
While \cite{du2018delay} also compared randomized networks and reported bidirectionality in the delay propagation, our analysis utilized the directed configuration model \citep{newman2001random} to generate randomized networks with the same degree distributions. 
As a result, we found that when the influence of degree distribution is removed, the delay propagation tends to be unidirectional. 
Moreover, we found that this property is universal regardless of the amount of delays and the airlines.
This result is contrary to the conclusions of \cite{du2018delay}.
However, it should be noted that our findings may not necessarily generalize to other air transport systems.
Furthermore, we discovered that the specific network motifs appear more (or less) frequently in the delay causality networks than the corresponding randomized counterparts.
These results suggest that delays tend to propagate in a unidirectional way. 
The results also suggest that the delays spread to a third airport from both airports when bidirectional propagation occurs within a pair of airports. 
The network motifs that tend to appear less frequently than in the randomized networks exhibit patterns opposite to these tendencies, supporting our conclusions. 
We also revealed that this characteristic is more profound in the groups with more significant delays. 
These findings suggest that delays tend to propagate following specific directional patterns, which could significantly contribute to predicting air traffic delays.
\add{For example, real-time monitoring of the delay-indicative network motifs that have been identified in this study could enhance the accuracy of predicting delay propagation. In the context of the Japanese aviation system, while previous research has focused on delays at the airport level, particularly at hub airports like Haneda, our findings on network motifs offer a completely new perspective on delay propagation at the network level.}

Additionally, our analysis unveiled airline-specific delay propagation patterns.
First, we observed that while the flight networks of ANA and JAL are relatively similar throughout the year, their delay causality networks differ significantly.
The central airports of spreading delays also differ greatly depending on the groups for ANA and JAL.
In the case of ANA, we observed some seasonality in the occurrence of delay propagation patterns.
Group 2 of ANA frequently appears in the summer, and the central airports of spreading delays are mainly located in Okinawa and Kyusyu islands.
Strong typhoons frequently hit these regions in the summer compared to the rest of the country, which explains this result.
In contrast, group 3 of ANA frequently appears in the winter, and the large airports across the country are the central airports of spreading delays. 
On the other hand, we did not observe noticeable seasonal characteristics in the groups for JAL compared to ANA.
One possible reason is that ANA and JAL set different buffers at each airport \citep{hirata2018flight}, which may significantly impact the delay propagation patterns.
\add{Another possibility is that they have different approaches to using alternative aircraft for delay mitigation.}
Identifying the causes of the observed differences between ANA and JAL warrants future work.

There are many other interesting future directions.
First, we only examined one year, 2019, but it would be interesting to investigate the delay propagation patterns over a more extended period.
The structure of air transport networks can change drastically in response to socio-economic events or airlines' strategic decisions \citep{sugishita2021recurrence}.
Upon these events, the delay propagation patterns could also change significantly. 
For example, the COVID-19 pandemic had a significant impact on the aviation industry. 
The delay propagation patterns may have changed considerably, and it is worth analyzing whether the results obtained in this study would still hold true in the post-pandemic period.
Additionally, factors such as the construction of airport runways and changes in air traffic control could significantly impact the overall delay propagation structure of the network. 
Our framework can be used to investigate these aspects as well.
Second, while we analyzed only two largest airlines, ANA and JAL,  future research should address the delay propagation patterns of smaller airlines. This could illuminate differences in delay propagation patterns between full-service carriers and low-cost carriers.
Third, it would be worth investigating the interplay between community structure and delay propagation, which has not been fully explored \citep{rocha2017dynamics}. 
The Japanese air transport network does not have a clear community structure.
Therefore, studying the recurrence of air transport networks in other countries or regions warrants future work. 
In air transport networks with community structures, more obvious recurrent patterns might be discovered if delays mainly propagate within the communities.
Fourth, investigating delay propagation on a finer time scale within a day would be desirable. 
Even if delay propagation patterns appear non-recurrent when constructing the network daily, they might be more recurrent up to a specific time of day.
It is essential to uncover this point to enable real-time prediction of air traffic delays.
Fifth, it would be interesting to integrate our framework with alternative causality tests such as nonlinear Granger causality test \citep{jia2022delay}, mutual information \citep{hlavavckova2007causality}, transfer entropy \citep{xiao2020study}, and convergent cross mapping \citep{guo2022detecting}.
Sixth, controlling delay propagation is a critical challenge. While causality analyses can contribute to understanding the delay propagation, they are insufficient for directly proposing policies to control delays. For example, identified causal relationships may be influenced by unobserved confounding factors such as weather conditions. Therefore, properly interpreting causal relationships while considering factors like weather conditions, flight schedules, and air traffic control, and utilizing them to establish effective control strategies, remains a significant challenge in the future.

\section*{Declaration of competing interest}
The authors declare no competing financial or non-financial interests.

\section*{Acknowledgements}
We appreciate Japan Society for the Promotion of Science (KAKENHI Grant-in-Aid for Early-Career Scientists \#22K14438) and Japan Airlines Co., Ltd. for supporting our research.

\appendix 
\section{\add{Distance matrix and dendrogram of hierarchical clustering}}
\begin{figure}[ht!]
\centering
\includegraphics[width=\linewidth]{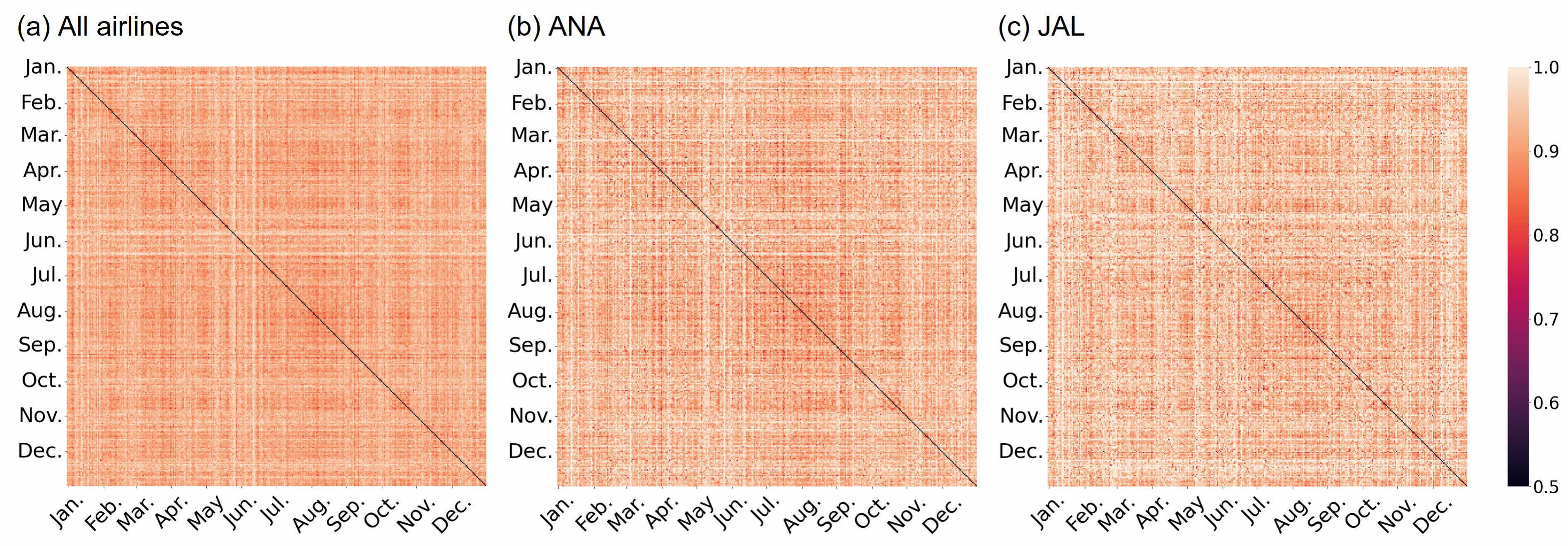}
\caption{Distance matrices of the delay causality networks for (a) all airlines, (b) ANA, and (c) JAL.
The colors indicate the network distance between pairs of the delay causality networks.
Note that the values of the diagonal elements are always zero, so the color scale is for the values except for the diagonal elements.}
\label{fig:distance_matrix}
\end{figure}

\begin{figure}[ht!]
\centering
\includegraphics[width=\linewidth]{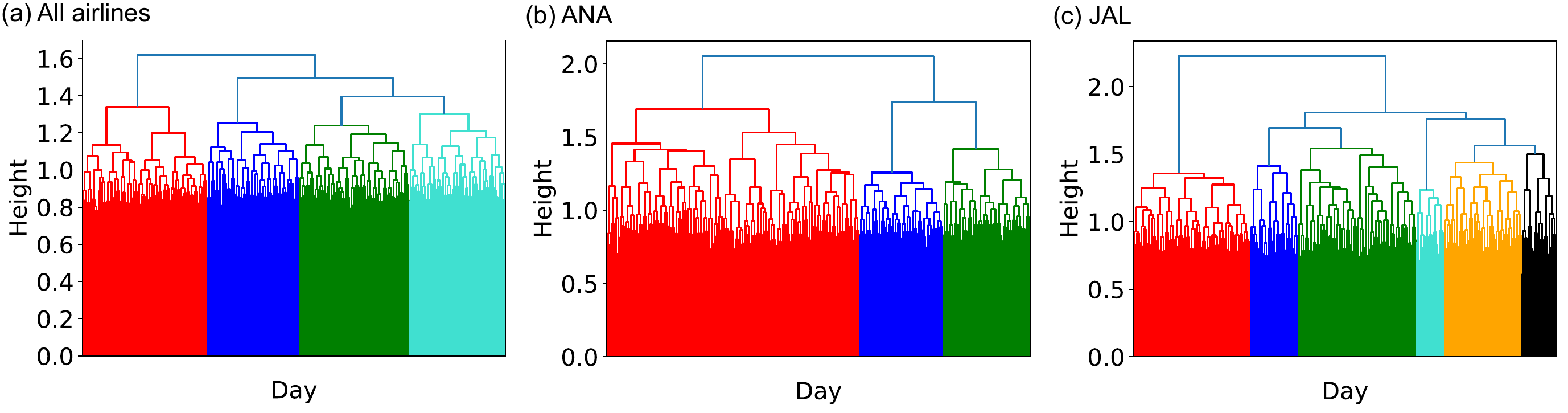}
\caption{Dendrogram for (a) all airlines, (b) ANA, and (c) JAL.
The colors indicate the groups that are identified by the hierarchical clustering. 
The number of groups is determined by optimizing Dunn's index.
}
\label{fig:dendrogram}
\end{figure}

\add{We show the distance matrices for all airlines, ANA, and JAL in 2019 in Fig.~\ref{fig:distance_matrix}. 
The colors indicate the values of network distance $L$. 
Higher values indicate more dissimilar networks. 
We cannot observe any clear change points or periodicity from the visual inspection of the distance matrices. 
We then show the dendrogram based on the hierarchical clustering in Fig.~\ref{fig:dendrogram}.
The number of groups is determined by maximizing Dunn's index. 
As a result, we identified four, three, and six groups for all airlines, ANA, and JAL, respectively.}

\section{Network motif for ANA and JAL}
\begin{figure}[H]
\centering
\includegraphics[width=0.64\linewidth]{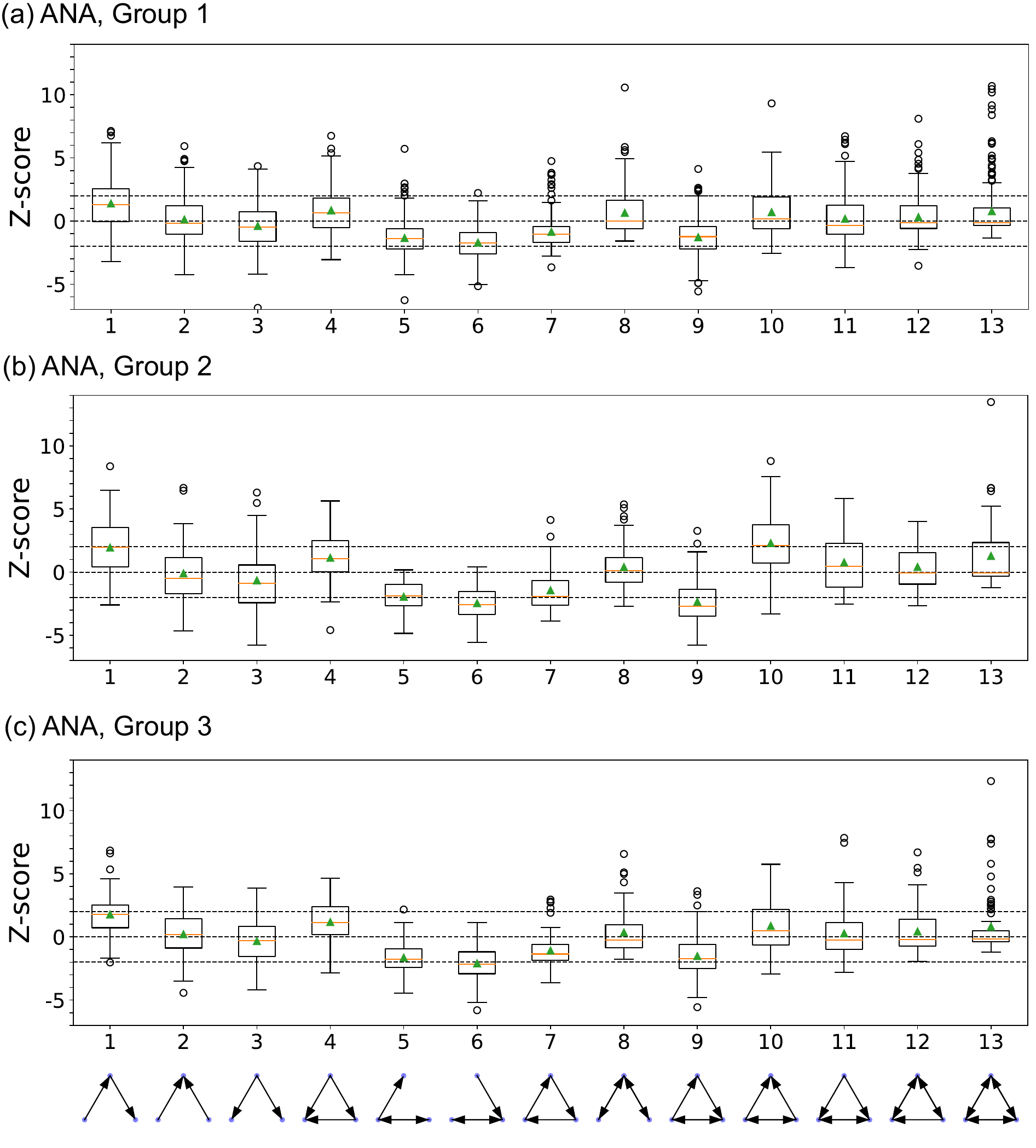}
\caption{Z-score of the network motifs for each group of ANA.
}
\label{fig:motif_ANA}
\end{figure}

\begin{figure}[H]
\centering
\includegraphics[width=0.64\linewidth]{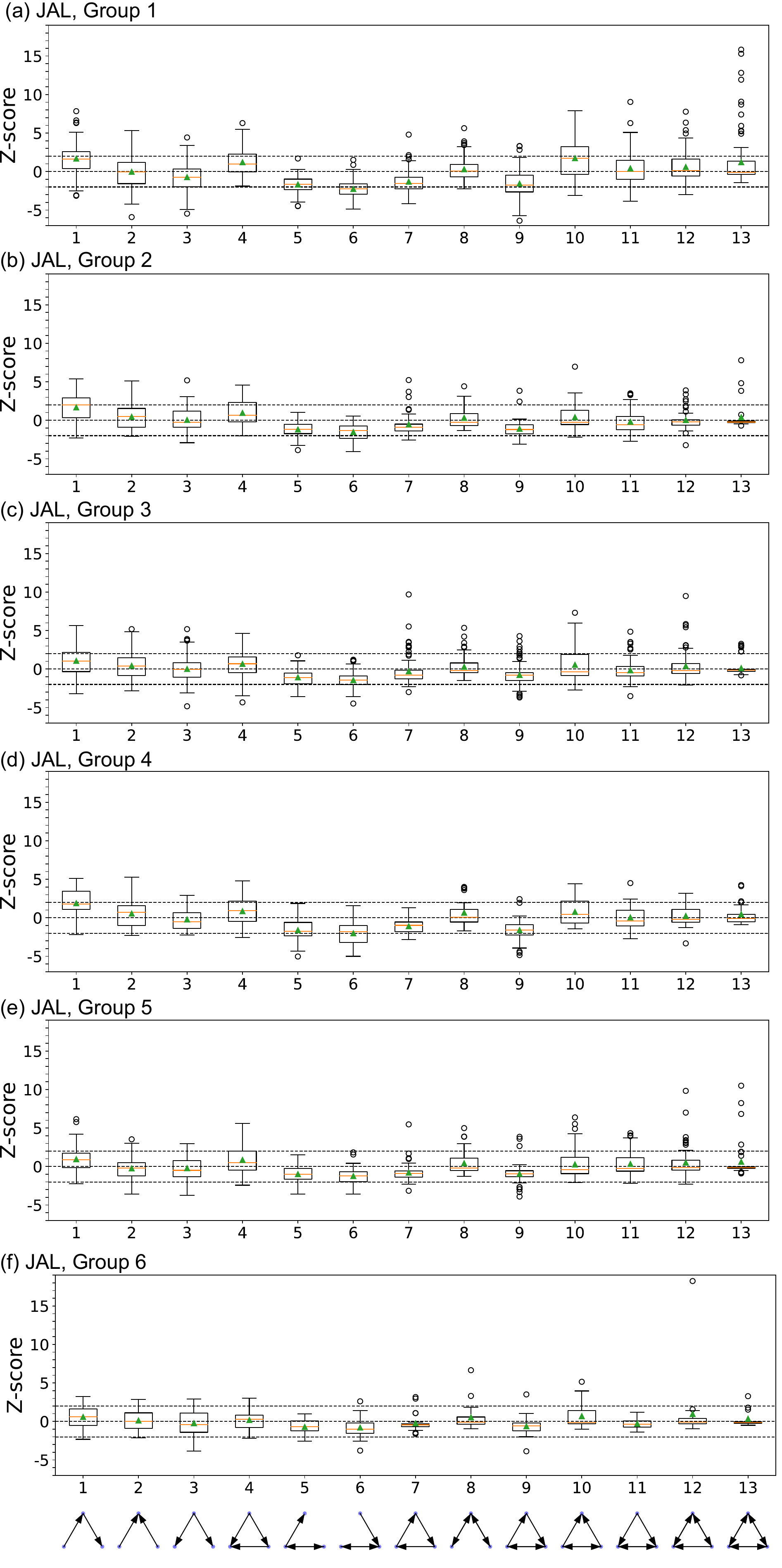}
\caption{Z-score of the network motifs for each group of JAL.
}
\label{fig:motif_JAL}
\end{figure}

We show the z-score distributions of the network motifs for each group of ANA and JAL in Figs.~\ref{fig:motif_ANA} and \ref{fig:motif_JAL}, respectively.

\bibliographystyle{elsarticle-harv}\bibliography{ref}
\end{document}